\documentstyle[aps,prb,epsf]{revtex}

\begin{document}

\draft


\title{Symmetry Origin of the Phase Transitions and Phase Separation in Manganites at Low Doping}
\author{Fan Zhong and Z. D. Wang}
\address{Department of Physics, University of Hong Kong, Hong Kong, P. R. China}
\date{\today}
\maketitle

\begin{abstract}
We analyze the symmetry changes of the paramagnetic to the A-type antiferromagnetic
and to the ferromagnetic phase transitions in undoped and moderately doped
LaMnO$_3$, respectively. We show that in the orthorhombic--distorted perovskite
manganites the phase separation at low doping is associated with the
noncollinear nature of the magnetic orders permitted by symmetry.
A simple model for the competition between the two phase transitions is
put forward within the framework of Landau theory of phase transitions.
\end{abstract}

\vspace{0.2in}


\preprint{Phys. Rev. B59, Nov, 1999}


\widetext

The discovery of ``colossal'' magnetoresistance (CMR) has stimulated a
renaissance of interest in doped rare--earth manganese oxides because of
their promising practical applications and their similarity to the cuprate
superconductor. \cite{wollan,cmr} Although great efforts have been devoted
to this system, the various phase transitions occurring under doping is still
not fully understood as a result of the complex interplay among magnetic,
charge, orbital and structural orders. Present lack of precise command of
strong correlations makes difficult discriminating models based on, for
example, pure double exchange, Jahn--Teller and doping variants. \cite{theory}
Therefore, it is desirable to investigate such general properties as symmetry
of the system that are feasible and meanwhile informative enough both to
impose rigorous general restrictions and to shed light to microscopic theories.

The perovskite--structured LaMnO$_{3}$ is believed to be a Jahn--Teller
distorted orthorhombic structure with a crystallographic space group $%
Pnma(D_{2h}^{16})$ at room temperatures. \cite{pickett} Below $T_{N} \sim 140$K,
it undergoes a magnetic transition from a paramagnetic (PM) to an $A$--type
antiferromagnetic (AFM) phase in which ferromagnetic (FM) layers are coupled
antiferromagnetically, different from the usual AFM couplings along all
nearest neighbors ($G$--type), while its lattice remains
unaltered. \cite{wollan} The Mn$^{3+}(d^{4})$ ion is believed to be in the
$t_{2g}^{3}e_{g}^{1}$ high--spin state; and strong on--site correlations
render the compound insulating in both magnetic phases. Upon doping of
divalent ions, some Mn ions lose their Jahn-Teller active $e_{g}^{1}$
electrons leaving much smaller Mn$^{4+}$ ions with mobile holes, a
sufficient amount of which may make the low--temperature phase metallic and
ferromagnetic via a double exchange, \cite{de} a superexchange interaction
between localized $t_{2g}$ spins which is facilitated by an external magnetic
field and hence follows the so--called colossal magnetoresistance. In
addition to this magneto-transport behavior, many unusual phenomena show
up such as the magneto--structural transition, charge and orbital orders and
their stability to external influences such as magnetic field, pressure,
x--ray, electric field and light irradiations. \cite{imada} A particular
issue that pose a great challenge to theorists, \cite{nagaev} besides the
mechanism of CMR itself, is the tendency of the system to phase separation
not only at high doping through a first order FM to charge--ordered AFM
transition, \cite{radaelli} but also at low doping.
\cite{wollan,allodib,hennion} We shall show below by symmetry analysis that
both the PM to AFM and the PM to FM transition are induced by the same
irreducible corepresentation, which, among others, permits a common FM
component for both the AFM and FM phase. The competition between these two
phases upon doping leads to the electronic phase separation at low doping
when combined with the microscopic mechanism of double exchange.
 
First we analyze the symmetry change of the PM to AFM phase transition in
undoped LaMnO$_3$. Since the crystallographic space group remains unaltered
during the magnetic phase transition, this transition must associate with a
one--dimensional irreducible corepresentation of the magnetic group at the
center $({\bf k}=0)$ of the orthorhombic Brillouin zone. As a result, the PM
to AFM transition is governed by a single order parameter that acquires a
nonzero value representing the staggered magnetization below $T_{N}$.

The magnetic group of the PM phase contains the time--reversal operation
${\cal R}$ itself as one of its elements and so is the grey group
$Pnma1^{\prime}$, a direct product group of $Pnma$ and the group
$\{E, {\cal R}\}$ with $E$ being the identity operation. \cite{bradley}
All its irreducible corepresentations are (ICR) thus multiplied into a
doubled set of even and odd groups. Only the odd representations are
relevant as ${\cal R}$ reverse the direction of a spin; and so the
transition from a PM state can be described simply by the axial vector
basis functions of the irreducible representations of the space
group $Pnma$. \cite{toledano} Designate a space group element
by $\{R|{\bf t}_{R}+{\bf t}\}$, where $R$ represents a proper or improper
rotation, ${\bf t}_{R}$ a nonprimitive (fractional) translation associated
with $R$, and ${\bf t}$ a primitive translation, the eight coset
representatives of $Pnma$ with respect to the subgroup of pure
translations $\{E|{\bf t}\}$ are, \cite{kovalev}
\[
\begin{tabular}{cccccccc}
$h_{1}$ & $h_{2}$ & $h_{3}$ & $h_{4}$ & $h_{25}$ & $h_{26}$ & $h_{27}$ & $%
h_{28}$ \\ \hline
$\{ E|000\} $ & $\{ U_{x}|\frac{1}{2}\frac{1}{2}\frac{1}{2}\} $%
 & $\{ U_{y}|0\frac{1}{2}0\} $ & $\{ U_{z}|\frac{1%
}{2}0\frac{1}{2}\} $ & $\{ I|000\} $ & $\{ \sigma _{x}|%
\frac{1}{2}\frac{1}{2}\frac{1}{2}\} $ & $\{ \sigma _{y}|0\frac{1}{%
2}0\} $ & $\{ \sigma _{z}|\frac{1}{2}0\frac{1}{2}\} $%
\end{tabular}
\]
where $I$ denotes an inversion, $U_{x}$ a rotation by $\pi$ about $x$--axis
and $\sigma_{x}$ a reflexion about the plane perpendicular to $x$, etc, and
$h_{i}$'s are Kovalev's symbols. Then according to the character table of
the point group $D_{2h}$, Table \ref{chara},
the magnetic symmetries of the phases that arise from the corresponding
irreducible representations can be determined as follows:
\begin{equation}
\begin{tabular}{cccccccc}
$\tau ^{1}$ & $\tau ^{2}$ & $\tau ^{3}$ & $\tau ^{4}$ & $\tau ^{5}$ & $\tau
^{6}$ & $\tau ^{7}$ & $\tau ^{8}$ \\ \hline
$Pnma$ & $Pn'm'a'$ & $Pnm'a'$ & $Pn'ma$ & $Pn'ma'$ & $Pnm'a$ & $Pn'm'a$ & $Pnma'$
\end{tabular}
.\label{symm}
\end{equation}
Here the primes indicate the symmetry elements that are associated with
${\cal R}$ in the respective magnetic groups, for example, all the reflexion
planes of $\tau^2$ must combine with ${\cal R}$ to give $Pn'm'a'$ since all
their corresponding characters are $-1$.\cite{bertaut}

\begin{minipage}{42.5pc}
\begin{minipage}{20.5pc}
\begin{table}
\caption{Characters of the irreducible representations of $Pnma$ at
${\bf k}=0$}
\label{chara}
\begin{tabular}{c|rrrrrrrr}
& $h_{1}$ & $h_{2}$ & $h_{3}$ & $h_{4}$ & $h_{25}$ & $h_{26}$ & $h_{27}$ & $%
h_{28}$ \\ \hline
$\tau ^{1}$ & $1$ & $1$ & $1$ & $1$ & $1$ & $1$ & $1$ & $1$ \\ 
$\tau ^{2}$ & $1$ & $1$ & $1$ & $1$ & $-1$ & $-1$ & $-1$ & $-1$ \\ 
$\tau ^{3}$ & $1$ & $1$ & $-1$ & $-1$ & $1$ & $1$ & $-1$ & $-1$ \\ 
$\tau ^{4}$ & $1$ & $1$ & $-1$ & $-1$ & $-1$ & $-1$ & $1$ & $1$ \\ 
$\tau ^{5}$ & $1$ & $-1$ & $1$ & $-1$ & $1$ & $-1$ & $1$ & $-1$ \\ 
$\tau ^{6}$ & $1$ & $-1$ & $1$ & $-1$ & $-1$ & $1$ & $-1$ & $1$ \\ 
$\tau ^{7}$ & $1$ & $-1$ & $-1$ & $1$ & $1$ & $-1$ & $-1$ & $1$ \\ 
$\tau ^{8}$ & $1$ & $-1$ & $-1$ & $1$ & $-1$ & $1$ & $1$ & $-1$%
\end{tabular}
\end{table}
\end{minipage}
\begin{minipage}{19pc}
\begin{figure}
\epsfysize 2in 
\epsfbox{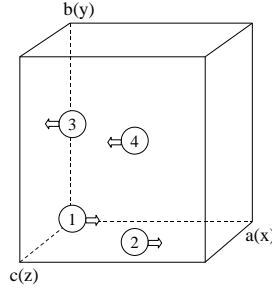}
\caption{Elementary unit-cell containing four Mn ions with magnetic moments shown schematically by arrows.}
\label{unit}
\end{figure}
\end{minipage}
\begin{minipage}{20.5pc}
\begin{table}
\caption{Transformation table of the magnetic moments}
\label{table1}
\begin{tabular}{c|rrrrrrrr}
& $h_{1}$ & $h_{2}$ & $h_{3}$ & $h_{4}$ & $h_{25}$ & $h_{26}$ & $h_{27}$ & $%
h_{28}$ \\ \hline
$\mbox{{\boldmath $\mu$}}_{1}$ & $\mbox{{\boldmath $\mu$}}_{1}$ & $\mbox{{\boldmath $\mu$}}_{4}$ & $\mbox{{\boldmath $\mu$}}_{3}$ & $%
\mbox{{\boldmath $\mu$}}_{2}$ & $\mbox{{\boldmath $\mu$}}_{1}$ & $\mbox{{\boldmath $\mu$}}_{4}$ & $\mbox{{\boldmath $\mu$}}_{3}$ & $%
\mbox{{\boldmath $\mu$}}_{2}$ \\ 
$\mbox{{\boldmath $\mu$}}_{2}$ & $\mbox{{\boldmath $\mu$}}_{2}$ & $\mbox{{\boldmath $\mu$}}_{3}$ & $\mbox{{\boldmath $\mu$}}_{4}$ & $%
\mbox{{\boldmath $\mu$}}_{1}$ & $\mbox{{\boldmath $\mu$}}_{2}$ & $\mbox{{\boldmath $\mu$}}_{3}$ & $\mbox{{\boldmath $\mu$}}_{4}$ & $%
\mbox{{\boldmath $\mu$}}_{1}$ \\ 
$\mbox{{\boldmath $\mu$}}_{3}$ & $\mbox{{\boldmath $\mu$}}_{3}$ & $\mbox{{\boldmath $\mu$}}_{2}$ & $\mbox{{\boldmath $\mu$}}_{1}$ & $%
\mbox{{\boldmath $\mu$}}_{4}$ & $\mbox{{\boldmath $\mu$}}_{3}$ & $\mbox{{\boldmath $\mu$}}_{2}$ & $\mbox{{\boldmath $\mu$}}_{1}$ & $%
\mbox{{\boldmath $\mu$}}_{4}$ \\ 
$\mbox{{\boldmath $\mu$}}_{4}$ & $\mbox{{\boldmath $\mu$}}_{4}$ & $\mbox{{\boldmath $\mu$}}_{1}$ & $\mbox{{\boldmath $\mu$}}_{2}$ & $%
\mbox{{\boldmath $\mu$}}_{3}$ & $\mbox{{\boldmath $\mu$}}_{4}$ & $\mbox{{\boldmath $\mu$}}_{1}$ & $\mbox{{\boldmath $\mu$}}_{2}$ & $%
\mbox{{\boldmath $\mu$}}_{3}$%
\end{tabular}
\end{table}
\end{minipage}
\end{minipage}

Next we determine the nature of the magnetic order below $T_{N}$. The
elementary unit--cell of LaMnO$_{3}$ contains four formula units with
Mn$^{3+}$ ions located at the $4a$ sites $1(000), 2(\frac{1}{2}0\frac{1}{2}),
3(0\frac{1}{2}0)$, and $4(\frac{1}{2}\frac{1}{2}\frac{1}{2})$
(see Fig.\ref{unit}) \cite{inter}. Associating each ion with a magnetic
moment {\boldmath $\mu$}, one can find their transformations by $h_{i}$'s as
shown in Table \ref{table1}. Let
\begin{eqnarray}
{\bf M} &=&\mbox{{\boldmath $\mu$}}_{1}+\mbox{{\boldmath $\mu$}}_{2}+\mbox{{\boldmath $\mu$}}_{3}+\mbox{{\boldmath $\mu$}}_{4} 
\nonumber \\
{\bf L}_{1} &=&\mbox{{\boldmath $\mu$}}_{1}-\mbox{{\boldmath $\mu$}}_{2}+\mbox{{\boldmath $\mu$}}_{3}-\mbox{{\boldmath $\mu$}}_{4} 
\nonumber \\
{\bf L}_{2} &=&\mbox{{\boldmath $\mu$}}_{1}+\mbox{{\boldmath $\mu$}}_{2}-\mbox{{\boldmath $\mu$}}_{3}-\mbox{{\boldmath $\mu$}}_{4} 
\nonumber \\
{\bf L}_{3} &=&\mbox{{\boldmath $\mu$}}_{1}-\mbox{{\boldmath $\mu$}}_{2}-\mbox{{\boldmath $\mu$}}_{3}+\mbox{{\boldmath $\mu$}}_{4},
\end{eqnarray}
which represent, respectively, the total magnetization and three possible AFM
collinear orders of $C$, $A$ and $G$ types, \cite{cmr,bertaut,toledano} then
according to Table \ref{table1}, the transformation properties of ${\bf M}$
and ${\bf L}_i$'s can be derived. Noting that both $U_x$ and $\sigma_x$
change the sign of the $y$ and $z$ components of an axial vector, one deduces
further the transformation properties of their respective components, from w
hich those components that form bases of the irreducible corepresentations
(ICR) of $Pnma1^{\prime }$ at ${\bf k}=0 $\ can be found to be
\begin{equation}
\begin{tabular}{cc}
ICR & \quad BASES \nonumber\\ \hline
$\tau ^{1}$ &\quad $L_{3x},L_{1y},L_{2z}$ \nonumber\\
$\tau ^{3}$ &\quad $M_{x},L_{2y},L_{1z}$ \nonumber\\
$\tau ^{5}$ &\quad $L_{2x},M_{y},L_{3z}$ \nonumber\\
$\tau ^{7}$ &\quad $L_{1x},L_{3y},M_{z}$,
\end{tabular}
\label{basis}
\end{equation}
that is, $L_{2x}$, for instance, transforms according to the representation
$\tau^5$, so do $M_y$ and $L_{1z}$.
Accordingly, it is straightforward to construct the magnetic Landau free--energy that
is an invariant:
\begin{eqnarray}
F &=&\sum_{i=1}^{3}\frac{a_{i}}{2}{\bf L}_{i}^{2}+\frac{c}{2}{\bf M}^{2}+\sum_{i=1}^{3}\frac{b_{i}}{4}{\bf L}_{i}^{4}+\frac{d}{4}{\bf M}^{4} 
+\frac{1}{2}\sum_{\alpha =x,y,z}(\sum_{i=1}^{3}\nu _{i\alpha }L_{i\alpha
}^{2}+\beta _{\alpha }M_{\alpha }^{2})  
+\gamma _{1}L_{3x}L_{1y}+\gamma _{2}L_{3x}L_{2z}+\gamma _{3}L_{1y}L_{2z} + \nonumber \\
&& \gamma _{4}M_{x}L_{2y} +\gamma _{5}M_{x}L_{1z}+\gamma _{6}L_{2y}L_{1z} 
+\gamma _{7}L_{2x}M_{y}+\gamma _{8}L_{2x}L_{3z}+\gamma _{9}M_{y}L_{3z} 
+\gamma _{10}L_{1x}L_{3y}+\gamma _{11}L_{1x}M_{z}+\gamma _{12}L_{3y}M_{z}.
\label{fall}
\end{eqnarray}
We have expanded the exchange contributions (first four terms) to the fourth order
and the magnetic anisotropic energies (the remaining terms) to the second order
because of their relatively smaller magnitude. Among all the coefficients,
$b_{i}$ and $d$ are positive for stability, and $\gamma_{i}$,
$\nu_{i\alpha}$, and $\beta_{\alpha}$ are small constants from relativistic
effects. \cite{toledano,dzyal} By ignoring the anisotropic contributions,
it is readily seen that Eq.(\ref{fall}) may yield FM or AFM phase of $G$,
$A$ or $C$ type depending on the coefficients $a_{i}$ and $c$.

Experimentally, it has been observed that the magnetic structure of
LaMnO$_3$ is $A$--type AFM order with the magnetic moments directing
primarily along $x$ axis. \cite{wollan,matsumoto} As a result, the magnetic
transition is described by a non--zero $L_{2x}$ below $T_{N}$.
So $a_{2}$ should become negative first among $a_{i}$ and $c$ upon cooling.
Retaining only those terms in Eq.(\ref{fall}) that contain the components
pertinent to ${\bf L}_{2x}$, we have
\begin{eqnarray}
F^{\prime } &=&\frac{a_{2}}{2}L_{2x}^{2}+\frac{a_{3}}{2}L_{3z}^{2}+
\frac{b_{2}}{4}L_{2x}^{4}
+\frac{\nu_{2x}}{2}L_{2x}^{2}+\frac{c}{2}M_{y}^{2}+\frac{1}{2}\beta_{y}M_{y}^{2}+ \gamma_{7}L_{2x}M_{y}
+\gamma _{8}L_{2x}L_{3z},  \label{fprime}
\end{eqnarray}
with a solution
\begin{mathletters}
\label{s1}
\begin{eqnarray}
L_{2x}^{2} &=& -\frac{1}{b_{2}}(a_{2}+\nu _{2x}-\frac{\gamma _{7}^{2}}{c
+\beta_{y}}-\frac{\gamma _{8}^{2}}{a_{3}})\approx -\frac{a_{2}}{b_{2}}, \\ 
M_{y} &=& -\frac{\gamma _{7}}{c+\beta _{y}}L_{2x}\approx -\frac{\gamma
_{7}}{c}%
\sqrt{-\frac{a_{2}}{b_{2}}}, \\ 
L_{3z} &=& -\frac{\gamma _{8}}{a_{3}}L_{2x}\approx -\frac{\gamma _{8}}{a_{3}}%
\sqrt{-\frac{a_{2}}{b_{2}}},
\end{eqnarray}
\end{mathletters}
that minimizes $F'$, where the last approximate equalities in each line
neglect those terms that are orders of magnitude smaller.

Equations (\ref{s1}) corresponds to an $A$--type AFM order with the magnetic
moments directing primarily along $\pm x$--axis in alternative Mn--O layers
perpendicular to $y$--axis. Meanwhile, all the moments tilt slightly along
both $y$--axis and $z$--axis giving rise to a weak FM and a weak $G$--type
AFM order respectively in those two directions. The solution associates with
the irreducible corepresentations $\tau ^{5}$, and the magnetic group of
the asymmetry phase is thus $Pn'ma'$ from the lists in (\ref{basis}) and
(\ref{symm}). Actually, the orientation of the magnetic moments can also be
obtained directly from the irreducible representation except their relative
magnitude. In other words, all these three types of order along their
respective directions are simultaneously allowed by the symmetry. So
in the symmetry point of view, canted phase is allowed. A weak ferromagnetic
component along $y$ has been inferred and observed in experiments.
\cite{matsumoto} The magnetic structure obtained also agreeswith the results of the local--spin--density--approximation calculations.
\cite{solovyev} We note in passing that as the magnetic anisotropic
energies arise from the relativistic spin--orbit and spin--spin
interactions, the special arrangement of the magnetic moments implies a
corresponding ordering of the orientations of the orbital moments and spins
relative to the crystalline lattice.

We now move on to the effect of doping. An important feature of
Eqs.(\ref{s1}) is the global ferromagnetism along $y$ axis. Weak as it is,
the partial FM order in alternative $xz$ planes indicates that upon doping
this weak FM component should be so enhanced that the FM phase arising at
sufficient doping rates should also direct along this $y$ axis as observed
experimentally. \cite{wollan,matsumoto} In other words, the PM to FM phase
transition should also associate with $\tau^5$. Formally, this is induced
by the coupling of $L_{2x}$ with $M_y$ in Eq.(\ref{fall}). Microscopically,
the doped holes promote the mobility of the $e_g$ electrons that mediate
FM coupling. Accordingly, as doping increases, the FM coupling is enhanced
and hence $T_c$, the FM transition temperature increases. On the other hand,
doping suppresses the antiferromagnetism. Therefore, when the doping
level $\delta$ (we use here $\delta$ instead of $x$ to avoid confusion)
is not too large, we may assume that
\begin{eqnarray}
a_2(\delta) & = & a_0(T-T_N+a_0'\delta),\nonumber \\
c(\delta) & = & c_0(T-T_c-c_0'\delta),\label{a2cx}
\end{eqnarray}
where $T_N$ and $T_c$ denote the AFM and the FM transition temperatures
at $\delta=0$, respectively, and $a_0$, $a_0'$, $c_0$, and $c_0'$ are
positive constants. Then, once $\delta>\delta_c \equiv (T_N-T_c)/(a_0'-c_0')$,
the coefficient $c$ will become negative first upon cooling; and so the
system exhibits a PM to FM instead of AFM transition. In this case, similar
analysis yields a dominant magnetization $M \approx \sqrt{-c/d}$ with now
weak canting of $L_{2x} \approx -\gamma_7M/a_2$ and $L_{3z} \approx
-\gamma_9M/a_3$ in contrast to the AFM state. This simple approximation is
in qualitative agreement with the magnetic phase diagram at low doping as
in Refs. [\onlinecite{urushibara,schiffer}], namely, the AFM transition
temperature decreases but the FM one increases with increasing doping.

More importantly, this simple model for the competition between the two
phase transitions exhibits phase separations at low doping. It is possible
to extend the present theory to a generalized Ginzburg--Landau theory by
including Coulomb repulsion and Boltzmann entropy terms for the holes as
well as gradient contributions from spatial inhomogeneities to give a
quantitative account. Here to illustrate the essential point, it is
instructive to compare the bulk free--energy of a doped uniform AFM state
with that of a state composed of a hole--depleted AFM and a hole--rich FM
phase. To this end, note that a uniform ordered AFM state at a doping level
$\delta_0$ has a bulk free-energy $-a_2^2/4b_2$, neglecting the small
relativistic contribution. Accordingly, when it is separated into, for
instance, a pure AFM state with $\delta_L=0$ and another weaker AFM state
of a number fraction $n$ ($\delta_0 \leq n<1$) with a higher doping
$\delta_H=\delta_0/n$ due to the conservation of holes, its bulk
free--energy aloneis lowered by $(1-n)a_{0}^2 a_0'^2 \delta_0^2/4bn>0$. Meanwhile,
the FM component also gains a bulk free--energy $n c^2(\delta_H)/4d$ or
$(1-n)c_{0}^2 c_0'^2 \delta_0^2/4dn$ for the separation at high temperatures
when $c(\delta_0)>0$ but $c(\delta_H)<0$ or at low temperatures when
$c(\delta_L)<0$, respectively. While as intermediate temperatures, whether
or not the FM component alone favors a separation depends on the system
(via the parameters). These gained energies may possibly overtake those
cost for hole aggregation particularly for low doping levels at which the
aggregated holes can still be distant enough to reduce their Coulomb
repulsion. As a result, a doped system tends to separate into hole--rich
regions with the FM order and hole-depleted region with the AFM order.
In reality, these electronically separated regions may be broken into
microscopic pieces by the long range Coulomb interaction in order to spread
the charge uniformly. Furthermore, the FM and AFM orders may possess a
certain variable strength depending on the concentration of the doped holes owing to
their common FM component. This accounts qualitatively for the coexistence
of FM and AFM features \cite{wollan,allodib} and the liquid--like
distribution of FM droplets observed in neutron scattering experiments at
low doping. \cite{hennion}

Note that the symmetry relationship between the two phases plays an
essential role in the above analysis. The tendency to favor separation
is caused by an instability in the inverse compressibility $\sim
\partial^2 F/\partial \delta^2=-(\partial a_2/\partial \delta)^2/2b_2-a_2
(\partial^2a_2/\partial \delta^2)/2b_2<0$ for an AFM state since doped
holes always raise its energy by frustrating the AFM order. Similarly,
the FM state favors more holes to a certain extent in order to lower
kinetic energies through the double exchange. Nevertheless, for a
{\it single} AFM or FM state, a phase separation is hardly possible
because of the Coulomb repulsion for charged particles. In general,
this fact is taken into account within the Landau--Ginzburg theory by
including quadratic terms to assure stability. In the present case, however,
{\it both} the hole--depleted AFM phase and the hole--rich FM phase are
energetically favored relative to a uniform state. An underlying mechanism
that makes this separation feasibleis the close symmetry correlation that facilitates the transition from the
AFM state to the FM state by the itinerant holes. Thus, as the doped holes
hop, they enhance the weak FM component of the AFM state via the double--
exchange interactions. Such enhanced FM regions can catch more holes which
in turn can further strengthen the FM order. This avalanche effect due to a
common FM component could trigger the separation of the holes into hole--
depleted and hole--rich regions.

In conclusion, we have shown that both the PM to $A$--type AFM and the PM to
FM transition in undoped and moderately doped LaMnO$_3$, respectively,
correspond to the transition from a magnetic group $Pnma1'$'to $Pn'ma'$, and
associate with the irreducible corepresentation $\tau^5$ of the parent phase.
This irreducible representation allows an $A$ mode AFM, a $G$ mode AFM and
an FM order along the respective $a$, $c$ and $b$ axis in the $Pnma$ setting.
Accordingly, the $A$--type AFM phase of undoped LaMnO$_3$ also possesses an
FM component, albeit weak, which is identical in direction with the FM phase
present at moderate doping. This symmetry relation may lead to a phase
separation into hole--depleted AFM regions and hole--rich FM regions in
order to take energetic advantage of both phases. Such a competition of the
two relevant phases may also work in other systems like
La$_{2-2x}$Sr$_{1+2x}$Mn$_2$O$_7$ where phase separation was also observed,
\cite{perring} though their symmetry may be different. Nevertheless,
further investigation is awaited.

This work was supported by a URC fund and a CRCG grant of the University of
Hong Kong.

\end{document}